\documentclass[]{spie}  

\usepackage{amsmath,amsfonts,amssymb}
\usepackage{graphicx}
\usepackage[colorlinks=true, allcolors=blue]{hyperref}
\usepackage{multirow}
\usepackage{upgreek}
\usepackage{gensymb}
\usepackage{mathtools} 

\usepackage[english]{babel}
\usepackage{lastpage}
\usepackage{hyperref}
\usepackage{rotating} 
\usepackage{caption, tabularx}
\usepackage{enumitem, ragged2e}
\usepackage[svgnames]{xcolor}
\usepackage{csquotes}
\usepackage{appendix}
\usepackage{epstopdf}
\usepackage{cancel}
\usepackage{siunitx}
\title{Deconvolution in Fluorescence Lifetime imaging microscopy (FLIM)}  

\author{Varun Mannam$^*$}
\author{Xiaotong Yuan}
\author{Scott Howard}
\affil{Department of Electrical Engineering, University of Notre Dame, Notre Dame, IN 46556, USA}
\authorinfo{*e-mail: vmannam@nd.edu}

\begin{document} 
\maketitle

\begin{abstract}
Fluorescence lifetime imaging microscopy (FLIM) is an important technique to understand the chemical micro-environment in cells and tissues since it provides additional contrast compared to conventional fluorescence imaging. When two fluorophores within a diffraction limit are excited, the resulting emission leads to non-linear spatial distortion and localization effects in intensity (magnitude) and lifetime (phase) components. To address this issue, in this work, we provide a theoretical model for convolution in FLIM to describe how the resulting behavior differs from conventional fluorescence microscopy. We then present a Richardson-Lucy (RL) based deconvolution including total variation (TV) regularization method to correct for the distortions in FLIM measurements due to optical convolution, and experimentally demonstrate this FLIM deconvolution method on a multi-photon microscopy (MPM)-FLIM images of fluorescent-labeled fixed bovine pulmonary arterial endothelial (BPAE) cells.
\end{abstract}


\section{Introduction}
Fluorescence lifetime imaging microscopy (FLIM) is a powerful technique in biomedical research that provides enhanced molecular contrast as fluorescence lifetime \cite{FLIM2007} information in addition to the intensity information that is usually obtained from a conventional fluorescence microscopy \cite{FluorescenceMicroscopy}. Fluorescence microscopy illuminates the samples to excite fluorophores within cells and collect the emitted light, whereas, in FLIM, additional measurements estimate the intrinsic property of the fluorophore as fluorescence lifetime. In FLIM measurements, the lifetime information can be extracted using either TD-FLIM \cite{TCSPC, TimeGatingFLIM} that measures the time delay between excited and emitted pulsed laser using an exponential fit function or FD-FLIM \cite{SuperSensitivity,ProteinInteractions} that measures the relative change in magnitude and phase of the periodically modulated laser pulse. While FLIM has been leveraged in several biological applications, several crucial techniques from fluorescence microscopy has not been translated to FLIM. For example, deconvolution techniques used to compensate for system optical response (point spread function: PSF) \cite{zhao2021sparse} and perform denoising are commonplace (and often essential) steps in fluorescence microscopy but have not been rigorously applied to FLIM data. FLIM data extraction and interpretation is often a complicated and time-consuming process, as the raw FLIM data are multi-dimensional, and the FLIM algorithms are computationally intensive. Recently, machine learning (ML), particularly deep learning \cite{mannam2020machine}, has gained more attention for its great performance in various FLIM image processing tasks such as image denoising \cite{mannam2021convolutional, mannam2021real}, super-resolution \cite{deep-storm, mannam2021deep}, estimation of lifetime from raw FLIM data using convolutional neural networks (CNNs) \cite{netflics1, netflics2}, however, it requires training dataset \cite{mannam2020machine}. To overcome large training data problems, self-supervised machine learning methods are proposed, but they provide limited accuracy with huge computational time \cite{li2021reinforcing}. Also, unsupervised machine learning methods are proposed in the literature; however, these methods are useful to form lifetime clusters, and segmentation \cite{esposito2007, yide_phasors}. This paper presents the analytical, theoretical description of optical convolution in FLIM and how the resulting behavior is different from traditional fluorescence microscopy. We then present the theoretical grounding and experimental demonstration of FLIM deconvolution on multi-photon FLIM images of fluorescent-labeled fixed bovine pulmonary arterial endothelial (BPAE) cells captured using our custom-built setup \cite{zhang2021instant}.

\section{Methodology}
\begin{figure}[!ht]
\centering
\includegraphics[width=0.82\linewidth]{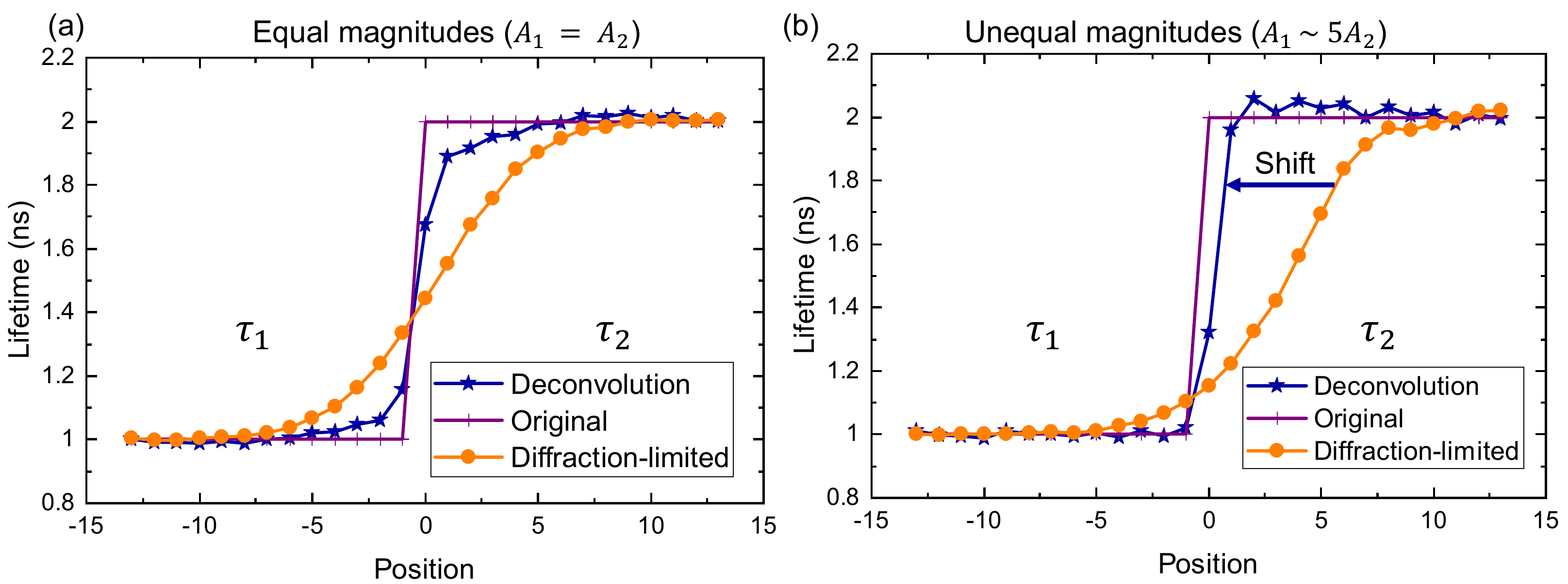}
\caption{Simulation results on the 1D data for two different fluorophores with different lifetime values. The effect of deconvolution of estimated lifetime in finding the ideal boundary between the two fluorophores with equal magnitudes (a) and unequal magnitudes (b).}\label{deconv_1D_res}
\end{figure}

When two fluorophores within a diffraction limit are excited in fluorescence microscopy, the resulting emission intensity in the convolution of PSF with emitters yields a linear response. In FLIM measurements, however the lifetime information convolution is highly non-linear. For example, in frequency domain FLIM, complex exponentials are summed together instead of purely real-values in fluorescence microscopy. Summation of complex exponentials in FLIM results in an effective blurring of the boundaries between two fluorophores (as seen in fluorescence microscopy) as well as a \textit{shifting} of the measured interface. This is represented in Fig.~\ref{deconv_1D_res}. Fig.~\ref{deconv_1D_res} shows this behavior from a simulated two-fluorophore model \cite{spie_overcoming} and the effect of convolution on the lifetime for two different scenarios. First, if the amplitude of two fluorophores are same as shown in Fig.~\ref{deconv_1D_res} (a)  and the second, if the magnitudes (magnitude is the product of concentration $C$ and cross-section $\sigma$ of a fluorophore) of the two-fluorophores are different as shown in Fig.~\ref{deconv_1D_res} (b). 

In Fig.~\ref{deconv_1D_res} (a), the magenta line shows the ideal lifetime of two fluorophores on a single 1D line where the left fluorophore has a lifetime of \SI{1}{ns} ($x<0$) and the right fluorophore has a lifetime of \SI{2}{ns} ($x>0$). The orange line indicates the estimated lifetime after applying system optical response (PSF) on this two-fluorophore model. Clearly, the orange line shows the effect of a smooth transition between lifetime values (missing ideal boundaries). One way to extract the ideal boundary is by applying a lifetime threshold (mean of two lifetime values). If the magnitudes of the fluorophores are equal, the threshold method can identify the ideal boundary between two fluorophores. However, this method doesn't work when the magnitudes are different, as illustrated in Fig.~\ref{deconv_1D_res} (b) orange curve. This is due to the unequal magnitudes leading to a shifting effect on the estimated lifetime. To solve this problem, we performed the Richardson-Lucy (RL) deconvolution \cite{dey2006richardson} on the 1D data using known PSF (used in the previous convolution step), and the results are shown in the blue line. Here RL method is an iterative approach for the maximum likelihood estimation for Poisson noise. Fig.~\ref{deconv_1D_res} (a) blue line indicates the estimated lifetime after applying deconvolution on the two-fluorophore FLIM data, which shows the ideal boundary. Similarly, Fig.~\ref{deconv_1D_res} (b) blue line corrects the shift (shift towards the ideal boundary) caused by the PSF convolution and can make it similar to the ideal boundary.

\textbf{Method}: The Richardson-Lucy (RL) deconvolution \cite{dey2006richardson} method applied in the previous step is explained here. Consider the estimated fluorophores is $o(s)$ at a position $s$ and PSF is $h(s)$. Measured FLIM data (either real or imaginary values of FLIM data or G and S of the phasor \cite{mannam2020machine}) follows the Poisson distribution, represented is $i(s)$. Next the original fluorophores real/imaginary values are estimated using the iterative approach using the measured values $i(s)$ and $h(s)$. 

\begin{equation}
    o_{k+1}(s) = \Bigg\{\frac{i(s)}{(o_k*h(s))} * h(-s) \Bigg\}.o_k(s) \label{RL_P}
\end{equation}
Here the ``$*$", indicates convolution operation, ``$.$" indicates point-wise multiplication, ``$k$" indicates the iteration index, respectively. However, the deconvolution operation leads to overshooting at the boundaries. The total-variation (TV) method is used as a regularization method to mitigate this issue. With the TV regularization, the final equation is modified as given below.
\begin{equation}
    o_{k+1}(s) = \Bigg\{\frac{i(s)}{(o_k*h(s))} * h(-s) \Bigg\}.\frac{o_k(s)}{1-\lambda div(\frac{\Delta o_k(s)}{|\Delta o_k(s)|})} \label{RL_P_TV}
\end{equation}

Here $\lambda$ indicates the regularization factor, typically small such as 0.005. The deconvolution operation, as shown in Eq.\ref{RL_P_TV} is applied on real and imaginary values of the measured FLIM data, and lifetime is extracted from the deconvolved real and imaginary FLIM measurements. If the PSF is unknown, blind deconvolution methods can estimate deconvolved results, including estimated PSF values. Recently, image processing tools such as ImageJ plugins are helpful to process the data, and the deconvolution can be applied to the FLIM data using the plugin \cite{sage2017deconvolutionlab2}. 

\section{Results and discussion}
In this section, the above-mentioned deconvolution method is extended to the 2D FLIM data and analysis.
\subsection{Simulated results}
\begin{figure}[!ht]
\centering
\includegraphics[width=0.95\linewidth]{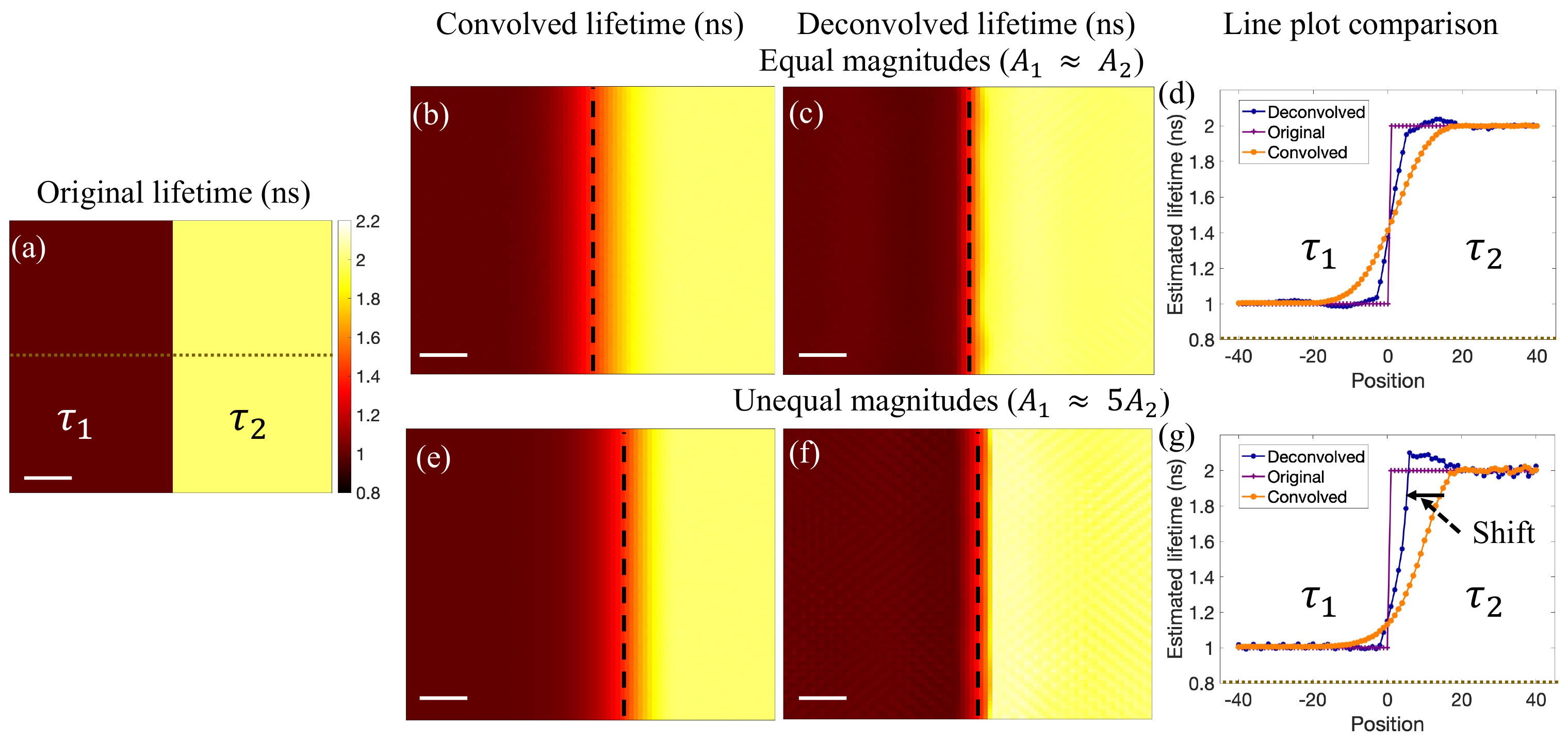}
\caption{Simulation results on the 2D data with two fluorophores with different lifetime values. (a) indicates the original objects with two fluorophore model. When the magnitude is similar in two fluorophores, convolved lifetime and deconvolved lifetime, line plots of the dotted line are given in (b), (c), and (d), respectively. Similarly, for the unequal magnitudes of two fluorophores, convolved lifetime and deconvolved lifetime, line plots of the dotted line are given in (e), (f), and (g), respectively. Scale bar: 1 $\mu$m.} \label{deconv_2D_res}
\end{figure}
Fig.~\ref{deconv_2D_res}(a) show the original lifetime of two-fluorophore model in frequency domain FLIM system with lifetime values of \SI{1}{ns} (left fluorophore: $\tau_1$) and \SI{2}{ns} (right fluorophore: $\tau_2$) respectively. In the first scenario, if the magnitude of the emitted signal is similar ($A_1 \approx A_2$) for each fluorophore, then the effect of convolution on lifetime is shown in Fig.~\ref{deconv_2D_res} (b). In the alternate scenario, consider that the magnitude of the emitted signal for one of the fluorophores is stronger compared to the other ($A_1 \approx 5 A_2$), and the corresponding lifetime after convolution with PSF is shown in Fig.~\ref{deconv_2D_res} (c). Clearly, in the second scenario, the ideal boundary between the two fluorophores is shifted towards the right by the stronger fluorophore, hence appears as inaccurate localization. Hence, the localization-based SR methods using lifetime information cannot be applied directly since it provides inaccurate results.

We demonstrate a RL-TV based deconvolution method that can correct the shift and identify the actual boundary between the two fluorophores to address this problem. First, we perform Richardson-Lucy (RL) deconvolution on each lifetime imaging component (phasor G and S, or real and imaginary). The RL deconvolution finds the maximum likelihood estimation of the original sample assuming Poisson noise. However, the real and imaginary (or phasor G and S) components do not follow Poisson statistics (more accurately follow a Skellam distribution \cite{griffin1992distribution}) and experimentally appear Gaussian-like. Similarly, the RL deconvolution method can be applied to the Gaussian noise distribution to extract the lifetime information. In addition, total variation (TV) based regularization is added to RL deconvolution to avoid additional artifacts, as commonly employed in fluorescence microscopy.

Fig.~\ref{deconv_2D_res} (c) and (d) shows the deconvolved lifetime when the magnitude of both fluorescence emission is equal. From this 1D data, deconvolved lifetime accurately separates two fluorophores along the intersection. Similarly, deconvolved lifetime for the unequal magnitudes scenario is shown in Fig.~\ref{deconv_2D_res} (f) and corresponding line plots are shown in Fig.~\ref{deconv_2D_res} (g). Clearly, from the unequal magnitudes scenario, the lifetime with deconvolution brings back the real boundary (shift towards the dominating fluorophore, see orange curve to blue curve). In Fig.~\ref{deconv_2D_res}, we show the convolution and deconvolution lifetime results after taking an average over 100 Gaussian noisy iterations with a standard deviation of $0.05$ (corresponding signal-to-noise ratio (SNR) is $26$ dB).

\subsection{Experimental results}
\begin{figure}[!ht]
\centering
\includegraphics[width=0.8\linewidth]{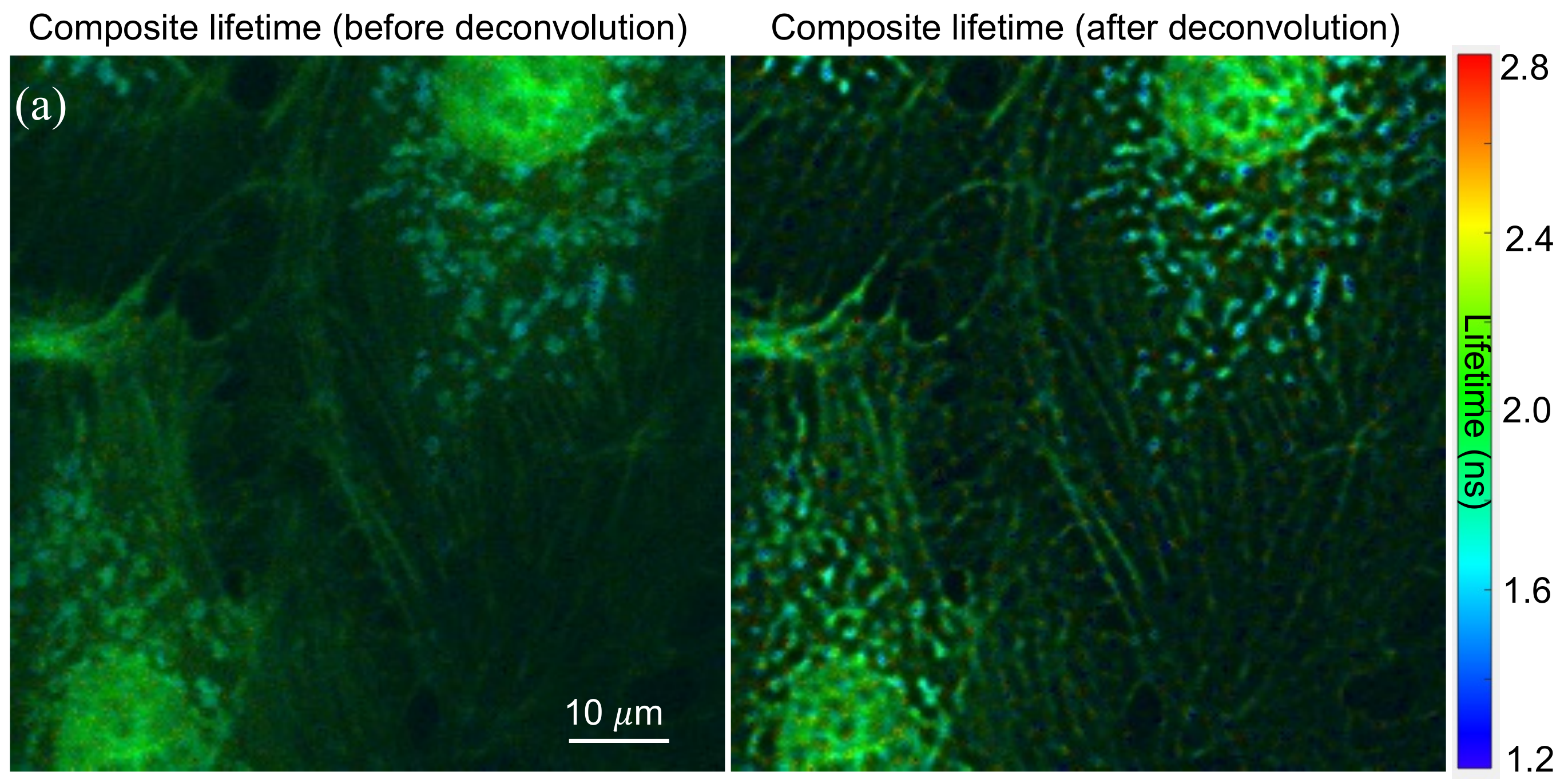}
\caption{Fixed BPAE sample composite lifetime before and after deconvolution on the lifetime information. Image acquired with custom two-photon FLIM microscope (see text). Scale bar: 10 $\mu$m.} \label{exp_results}
\end{figure}

The proposed deconvolution approach is demonstrated experimentally in Fig.~\ref{exp_results}. FLIM images of fixed BPAE cells [labeled with MitoTracker Red CMXRos (mitochondria), Alexa Fluor 488 phalloidin (F-actin), and DAPI (nuclei)] from Invitrogen FluoCells F36924 is presented before and after deconvolution. The lifetime images are captured with our custom-built two-photon FLIM setup \cite{zhang2021instant} with an excitation wavelength of 800 nm, single-channel (captured all the fluorophores simultaneously using a bandpass filter (300-700nm), an image dimension of 260 $\times$ 260, the pixel dwell time of 12 µs, total acquisition time is \SI{1}{s}, and pixel width of 300 nm.

Fig.~\ref{exp_results} (a) shows the experimentally captured FLIM lifetime information of BPAE cells with poor resolution. In contrast, after applying RL deconvolution on the real and imaginary values of FLIM data, the lifetime is extracted as shown in Fig.~\ref{exp_results} (b) shows a better resolution of nucleus and mitochondria fluorophores. The methods used in this manuscript are provided as open-source, accessible via GitHub\footnote{\url{https://github.com/ND-HowardGroup/Deconvolution_lifetime_imaging.git}}.

\section{Conclusions}
In Fluorescence lifetime imaging microscopy (FLIM), when two fluorophores within a diffraction limit are excited, the resulting emission leads to different spatial distortion and localization effects in intensity (magnitude) and lifetime (phase) components. To address this issue, in this work, we provide a theoretical model to demonstrate a Richardson-Lucy (RL) based deconvolution, including the TV-based regularization method to correct for the distortions in FLIM measurements due to optical convolution and provide an accurate boundary of the two-fluorophore model. The technique is experimentally demonstrated on MPM-FLIM images of fluorescent-labeled fixed bovine pulmonary arterial endothelial (BPAE) cells. Finally, this deconvolved lifetime information can effectively enhance biological structures of interest in biomedical imaging applications. 

\section*{Disclosures}
\noindent The authors declare no conflicts of interest.

\section*{Funding.}
This material is based upon work supported by the National Science Foundation (NSF) under Grant No. CBET-1554516. 

\acknowledgments 
The authors further acknowledge the Notre Dame Center for Research Computing (CRC) for providing the Nvidia GeForce GTX 1080-Ti GPU resources for performing the deconvolution on the fluorescence lifetime images and data analysis in MatLAB and Python.

\bibliography{report} 
\bibliographystyle{spiebib} 
\end{document}